\documentclass[pdflatex,aps,twocolumn,superscriptaddress,prl,fleqn,showpacs,nofootinbib,preprintnumbers]{revtex4}
\usepackage{amssymb,amsmath,epsfig}

\DeclareMathOperator{\tr}{Tr}

\renewcommand{\d}{\mathrm{d}}
\newcommand{\sT}{{\scriptscriptstyle T}}
\newcommand{\bm}[1]{\mbox{\boldmath $#1$}}

\begin{document}

\title{Linearly Polarized Gluons and the Higgs Transverse Momentum Distribution}

\author{Dani\"el Boer}
%\email{D.Boer@rug.nl}
\affiliation{Theory Group, KVI, University of Groningen,
Zernikelaan 25, NL-9747 AA Groningen, The Netherlands}

\author{Wilco J. den Dunnen}
%\email{w.den.dunnen@vu.nl}
\affiliation{Department of Physics and Astronomy, Vrije Universiteit
  Amsterdam, NL-1081 HV Amsterdam, The Netherlands}

\author{Cristian Pisano}
%\email{cristian.pisano@ca.infn.it}
\affiliation{Dipartimento di Fisica, Universit\`a di Cagliari, and INFN, Sezione di Cagliari, I-09042 Monserrato (CA), Italy}

\author{Marc Schlegel}
%\email{marc.schlegel@uni-tuebingen.de}
\affiliation{Institute for Theoretical Physics,
                Universit\"{a}t T\"{u}bingen,
                Auf der Morgenstelle 14,
                D-72076 T\"{u}bingen, Germany}

\author{Werner Vogelsang}
%\email{werner.vogelsang@uni-tuebingen.de}
\affiliation{Institute for Theoretical Physics,
                Universit\"{a}t T\"{u}bingen,
                Auf der Morgenstelle 14,
                D-72076 T\"{u}bingen, Germany}

\begin{abstract}
We study how gluons carrying linear polarization inside an unpolarized hadron 
contribute to the transverse momentum distribution of Higgs bosons produced
in hadronic collisions. They modify the distribution produced by unpolarized gluons
in a characteristic way that could be used to determine 
whether the Higgs boson is a scalar or a pseudoscalar particle.
\end{abstract}

\date{\today}

\maketitle

It is sometimes said that the LHC is a `gluon collider', because at high energies the gluon 
density inside a proton becomes dominant over the quark densities. Higgs production,
in particular, predominantly arises from gluon-gluon `fusion' $gg \to H$ 
through a triangular top quark loop. QCD corrections to this process have 
been calculated with increasing precision  
\cite{Georgi:1977gs,Dawson:1990zj,Djouadi:1991tka,Spira:1997dg,Catani:2001ic,Harlander:2001is,Anastasiou:2002yz}, 
making it well understood. It is not commonly 
known however that the LHC is actually also to 
some extent a {\it polarized} gluon collider, since gluons can in principle be linearly polarized 
inside an unpolarized proton. Their corresponding distribution, 
here denoted by $h_1^{\perp\, g}$ and first defined in 
Ref.\ \cite{Mulders:2000sh}, requires the gluons to have a nonzero transverse momentum
with respect to the parent hadron. It corresponds to an interference between 
$+1$ and $-1$ helicity gluon states that would be suppressed 
without transverse momentum. 

\begin{figure}[htb]
\centering
\psfig{file=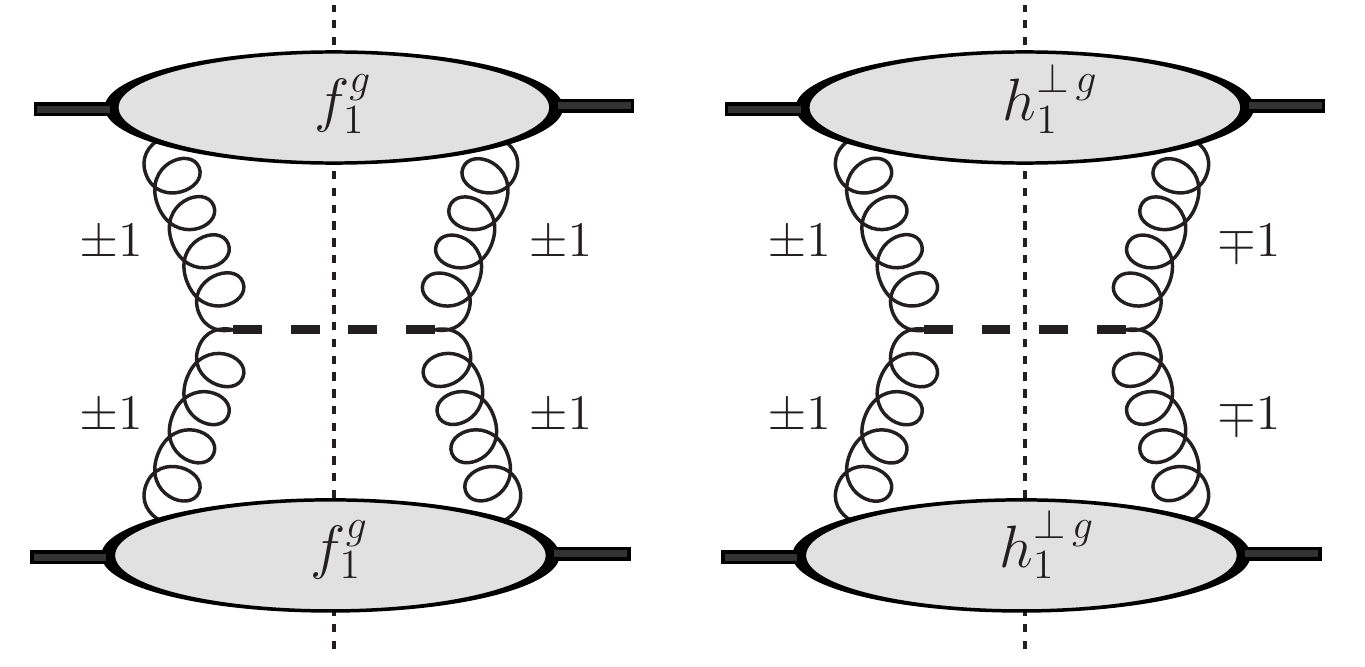, width=0.45\textwidth} %modDB
\caption{\it Gluon helicities in the $g g \to H$ squared amplitude for unpolarized (left) and linearly polarized 
production (right).}
\label{amplitudesquared}
\end{figure}

So far the function $h_1^{\perp\,  g}$ has not been studied experimentally, and consequently 
nothing is known about its magnitude. Only a theoretical upper bound has been given \cite{Mulders:2000sh,Boer:2010zf}. 
Recently, several ways of probing $h_1^{\perp\,  g}$ have been 
put forward, namely in heavy quark pair or dijet production \cite{Boer:2010zf}, or in 
photon pair production \cite{Qiu:2011ai}, where in all cases the transverse momentum of the pair is  
measured. One way in which linearly polarized gluons can manifest themselves in these processes
is through azimuthal asymmetries. However, it was found that they can also generate a term
in the cross section that is independent of  azimuthal angle. This happens when two linearly polarized 
gluons, one from each hadron, partici\-pate in the scattering. In this way they can also contribute to 
production of a scalar particle, such as a scalar or pseudoscalar 
Higgs boson, when its transverse momentum $q_\sT$ is measured. It has in fact been shown 
\cite{Nadolsky:2007ba,Catani:2010pd} that such a contribution is generated perturbatively.
In other words, if at tree level gluons are taken to be unpolarized, at order $\alpha_s$ they will become 
to some extent linearly polarized. In the transverse momentum distribution of spin-0 
particles produced in proton-proton collisions this will give rise to an additional 
contribution at order $\alpha_s^2$, because of the double helicity flip involved 
(see Fig.\ \ref{amplitudesquared}). While this may be expected to make only a relatively 
modest contribution, the function $h_1^{\perp\, g}$ is of {\it non-perturbative} nature and
is present at tree level already. Therefore, a significant influence of linearly polarized gluons 
on the distribution of the produced particle at low $q_\sT$ is not excluded. 

%%%%%%%%%%%%%%%%%%%%%%%%%%%%%%%%%%%%%%%%%%%%%%%%%%%%%%%%%%%%%%%%%%%%%%%%%%%%
%%%%%%%%%%%%%%%%%%%%%%%%%%%%%%%%%%%%%%%%%%%%%%%%%%%%%%%%%%%%%%%%%%%%%%%%%%%

In light of this, we will investigate in this letter how the distribution of linearly polarized gluons 
may affect the transverse momentum distribution of Higgs bosons for $q_\sT \ll m_H$, where $m_H$ 
is the Higgs mass. We shall observe that linearly polarized gluons 
may in fact provide a tool to uncover whether the Higgs boson is a scalar or a pseudoscalar particle.  Thus far, relatively few 
suggestions to this end 
have been put forward for the LHC, typically using azimuthal distributions, for example in 
Higgs + jet pair production \cite{Campanario:2010mi} or in $\tau$ pair decays \cite{Berge:2011ij}. 
The suggestion we put forward here does not involve measurements of any angular distributions. Instead,  
we will show that linear polarization of gluons simply leads to a modulation of the Higgs transverse momentum 
distribution that depends on the nature of the Higgs particle.  

Transverse momentum dependent distribution functions (TMDs) of gluons in an unpolarized
hadron are defined through a 
matrix element of a correlator of the gluon field strengths $F^{\mu \nu}(0)$ and
$F^{\nu \sigma}(\xi)$, evaluated at fixed light-front (LF) time
$\xi^+ =\xi{\cdot}n=0$, where $n$ is a lightlike vector
conjugate to the parent hadron's four-momentum $P$. Decomposing the gluon momentum as
$p = x\,P + p_\sT + p^- n$, the correlator is given by \cite{Mulders:2000sh}
\begin{eqnarray}
\label{GluonCorr}
\Phi_g^{\mu\nu}(x,\bm p_\sT )
& = &  \frac{n_\rho\,n_\sigma}{(p{\cdot}n)^2}
{\int}\frac{d(\xi{\cdot}P)\,d^2\xi_\sT}{(2\pi)^3}\
e^{ip\cdot\xi}\, \nonumber
\\
& & \hspace{-0.4 cm} \times
\langle P|\,\tr\big[\,F^{\mu\rho}(0)\,
F^{\nu\sigma}(\xi)\,\big]
\,|P \rangle\,\big\rfloor_{\text{LF}} \nonumber \\
&& \hspace{-2 cm} =
-\frac{1}{2x}\,\bigg \{g_\sT^{\mu\nu}\,f_1^g
-\bigg(\frac{p_\sT^\mu p_\sT^\nu}{M^2}\,
{+}\,g_\sT^{\mu\nu}\frac{\bm p_\sT^2}{2M^2}\bigg)
\;h_1^{\perp\,g} \bigg \} ,
\end{eqnarray}
with $p_{\sT}^2 = -\bm p_{\sT}^2$, $g^{\mu\nu}_{\sT} = g^{\mu\nu}
- P^{\mu}n^{\nu}/P{\cdot}n-n^{\mu}P^{\nu}/P{\cdot}n$, and $M$ the proton mass.
$f_1^g(x,\bm{p}_\sT^2)$ represents the
unpolarized gluon distribution and 
$h_1^{\perp\,g}(x,\bm{p}_\sT^2)$ the 
distribution of linearly polarized gluons. 
In (\ref{GluonCorr}) we have omitted a Wilson line that renders the correlator gauge invariant. As any TMD, $h_1^{\perp\, g}$ will receive 
contributions from initial and/or final state interactions, which make the gauge link process-dependent. 
Therefore, despite the fact that it is $T$-even, $h_1^{\perp\, g}$ can 
receive non-universal contributions, and its extraction can be hampered for processes where factorization does
not hold, such as dijet production in hadron-hadron collisions \cite{Boer:2009nc,Rogers:2010dm,Buffing:2011mj}. 
Higgs production, on the other hand, is expected to allow for TMD factorization, 
just like the Drell-Yan process. A more detailed study of this remains to be carried out.

The calculation of the Higgs production cross section in the TMD framework closely follows Refs.\ \cite{Boer:2007nd,Boer:2009nc}. The generic contribution by $gg\to H$ reads
\begin{eqnarray} %modDB
\hspace{0cm}\frac{E_{H}\,\d\sigma}{\d^{3}\vec{q}}\Big|_{q_{\sT}\ll m_{H}}  & =  & 
\frac{\pi x_a x_b}{16m_{H}^{2}S}\,\times \label{eq:CSTMDHiggs}\\
& &
\hspace{-3cm}\,\int d^{2}\bm p_{a\sT}\, \int d^{2}\bm p_{b\sT}\,\delta^{2}(\bm p_{a\sT}+
\bm p_{b\sT}-\bm q_{\sT})\,\Phi_g^{\mu \nu}(x_{a},\,\bm p_{a\sT})  \nonumber\\
& & \hspace{-3cm}\times \Phi_g^{\rho \sigma}(x_{b},\,\bm p_{b\sT})\, 
\left(\hat{\mathcal{M}}^{\mu \rho}\right)\left(\hat{\mathcal{M}}^{\nu \sigma}\right)^{\ast}
\Big|_{p_{a}=x_{a}P_{a}}^{p_{b}=x_{b}P_{b}}+\mathcal{O}\left(\frac{q_{\sT}}{m_{H}}\right).   \nonumber 
\end{eqnarray} 
For now, we assume on-shell production of the Higgs particle, with $\vec{q}$ and $E_H$
its momentum and energy. $P_a$ and $P_b$ are the momenta of the colliding protons, 
$S=(P_a+ P_b)^2$, and $x_{a(b)}=q^2/(2P_{a(b)}\cdot q)$.
To lowest order, the hard partonic amplitude $\hat{\mathcal{M}}$ is given by the well-known 
formula~\cite{Spira:1997dg} for the $gg\to H$ triangle diagram:
\begin{equation} %modDB
\hat{\mathcal{M}}_{H}^{\mu \nu} =  i 2^{1/4}G_F^{1/2} \alpha_s\,m_H^2\,g^{\mu \nu}_\sT
{\cal A}_{H}(\tau)/(8\pi)\label{eq:Scalar1}
\end{equation}
for a scalar Standard Model (SM)  Higgs boson $H^0$,
where we only consider top quarks in the triangle, and
where $G_F$ is the Fermi constant, $\alpha_s$ the strong coupling constant, 
$\tau= m_H^2/(4m_t^2)$ with the top mass $m_t$, and 
${\cal A}_{H}(\tau)=2\left(\tau + (\tau - 1)J(\tau)\right)/\tau^2$ with 
\begin{equation}
J(\tau)=\left\{ \begin{array}{c}
-\frac{1}{4}\left(\ln\left(\frac{1+\sqrt{1-1/\tau}}{1-\sqrt{1-1/\tau}}\right)-i\pi\right)^{2}\,,\, \tau>1\\[2mm]
\,\arcsin^{2}\left(\sqrt{\tau}\right)\,,\, \tau \le 1.\end{array}\right.\label{eq:VertInt}
\end{equation}
For a pseudoscalar Higgs boson $A^0$ with a simple coupling $g_t\,(\sqrt{2}G_F)^{1/2} m_t^2\,\gamma_5$ to quarks~\cite{Harlander:2002vv}
we have instead
\begin{equation} %modDB
\hat{\mathcal{M}}_{A}^{\mu \nu}  =  i 2^{1/4}G_F^{1/2}\alpha_{s}\,m_H^2\,\epsilon_{\sT}^{\mu \nu}\,
{\cal A}_{A}(\tau)/(8\pi),\label{eq:Pseudo1}
\end{equation}
where $\epsilon_{\sT}^{\mu \nu}$ is the two-dimensional Levi-Civita tensor, and ${\cal A}_{A}(\tau)=
g_t 2J(\tau)/\tau$. 
As mentioned above, QCD corrections to these amplitudes 
have been calculated. However, since our 
goal is to study the effect of linearly polarized gluons whose distribution $h_1^{\perp g}$ is anyway unknown, we limit 
ourselves to the lowest-order expressions (\ref{eq:Scalar1}) and (\ref{eq:Pseudo1}).
This leads to the 
following expressions for scalar and pseudoscalar Higgs production:
\begin{eqnarray}
\frac{E\,\d\sigma^{H(A)}}{\d^{3}\vec{q}}\Big|_{q_{\sT}\ll m_{H}} & = & \frac{\pi \sqrt{2}G_F}{128m_H^2 S}\left(\frac{\alpha_{s}}{4\pi}\right)^{2}\,\left|{\cal A}_{H(A)}(\tau)\right|^2 \nonumber\\
 & & \hspace{-2.7cm}\times \left(\mathcal{C}\left[f_{1}^{g}\, f_{1}^{g}\right]\pm\mathcal{C}\left[w_{H}\, h_{1}^{\perp g}\, h_{1}^{\perp g}\right]\right)\,+\mathcal{O}(\frac{q_{\sT}}{m_{H}})\,,\label{eq:CSscalarHiggs} 
 \end{eqnarray}
where the upper (lower) sign refers to the scalar (pseudoscalar) case, and where
we have the TMD convolution
\begin{eqnarray}
\mathcal{C}[w\, f\, f] & \equiv & \int d^{2}\bm p_{a\sT}\int d^{2}\bm p_{b\sT}\,
\delta^{2}(\bm p_{a\sT}+\bm p_{b\sT}-\bm q_{\sT})\nonumber\\
 & & \hspace{-0.5 cm} \times w(\bm p_{a\sT},\bm p_{b\sT})\, f(x_{a},\bm p_{a\sT}^{2})\, f(x_{b},\bm p_{b\sT}^{2})\,,\label{eq:Conv}
\end{eqnarray} 
with the transverse momentum weight
\begin{equation}
w_{H}=\frac{(\bm p_{a\sT}\cdot\bm p_{b\sT})^{2}-\frac{1}{2}\bm p_{a\sT}^{2}
\bm p_{b\sT}^{2}}{2M^{4}}\,.\label{eq:Higgsweight}
\end{equation}
We emphasize the sign difference in the 
$\mathcal{C}[w_{H}\, h_{1}^{\perp g}\, h_{1}^{\perp g}]$ term
in Eq.\ \eqref{eq:CSscalarHiggs}, 
which may offer an opportunity to determine the parity of the Higgs boson. 
The terms involving $h_{1}^{\perp g}$  have the model-independent
property $\langle \bm q_\sT^{2\alpha} \rangle_{hh} \equiv \int d^{2}\bm q_{\sT}\, (\bm q_\sT^{2})^\alpha\,
\mathcal{C}[w_{H}\, h_{1}^{\perp g}\, h_{1}^{\perp g}]=0$ for $\alpha=0,1$. 
This feature points towards a distinctive transverse momentum distribution of the 
$h_1 ^{\perp g} \, h_1 ^{\perp g}$ term with a double node in $\bm q_\sT$. 
We note that since $\langle 1 \rangle_{hh}=0$, linearly polarized gluons 
do not affect the $\bm q_\sT$-integrated cross section.

In the following we estimate the possible size of the contribution by linearly polarized gluons to the 
Higgs production cross section at tree level. 
Although the function $h_1^{\perp g}$ itself is unknown, a model-independent 
positivity bound for it has been derived in Ref.~\cite{Mulders:2000sh}:
\begin{equation}
\frac{\bm p_\sT^2}{2M^2}\,|h_1^{\perp g}(x,\bm p_\sT^2)|\le f_1^g(x,\bm p_\sT^2)\,,\label{eq:Bound}
\end{equation}
valid for all $x$ and $\bm p_\sT$. The maximally possible effect will be generated when this
bound is saturated. Models may also shed light on the size of $h_1^{\perp g}$. In the simple perturbative 
quark target model of gluon TMDs of Ref.~\cite{Meissner:2007rx} the function $h_1^{\perp g}$ is found 
to possess the same characteristic $1/x$ increase as the distribution of unpolarized gluons $f_1^g$, which 
suggests that linearly polarized gluons may be as relevant at small $x$ as unpolarized ones. Another recent 
model calculation~\cite{Metz:2011wb} shows saturation of the positi\-vi\-ty bound for $h_1^{\perp g}$ in 
heavy nuclei in certain transverse momentum regions (Weizs\"{a}cker-Williams model) or even over the 
full momentum range (dipole model). This suggests that saturation of the positivity bound at least locally 
in $x$ or $\bm p_\sT^2$ might not be an unrealistic assumption.

\begin{figure}[htb]
\vspace*{-2.2cm}
\hspace*{-4.7cm}
\psfig{file=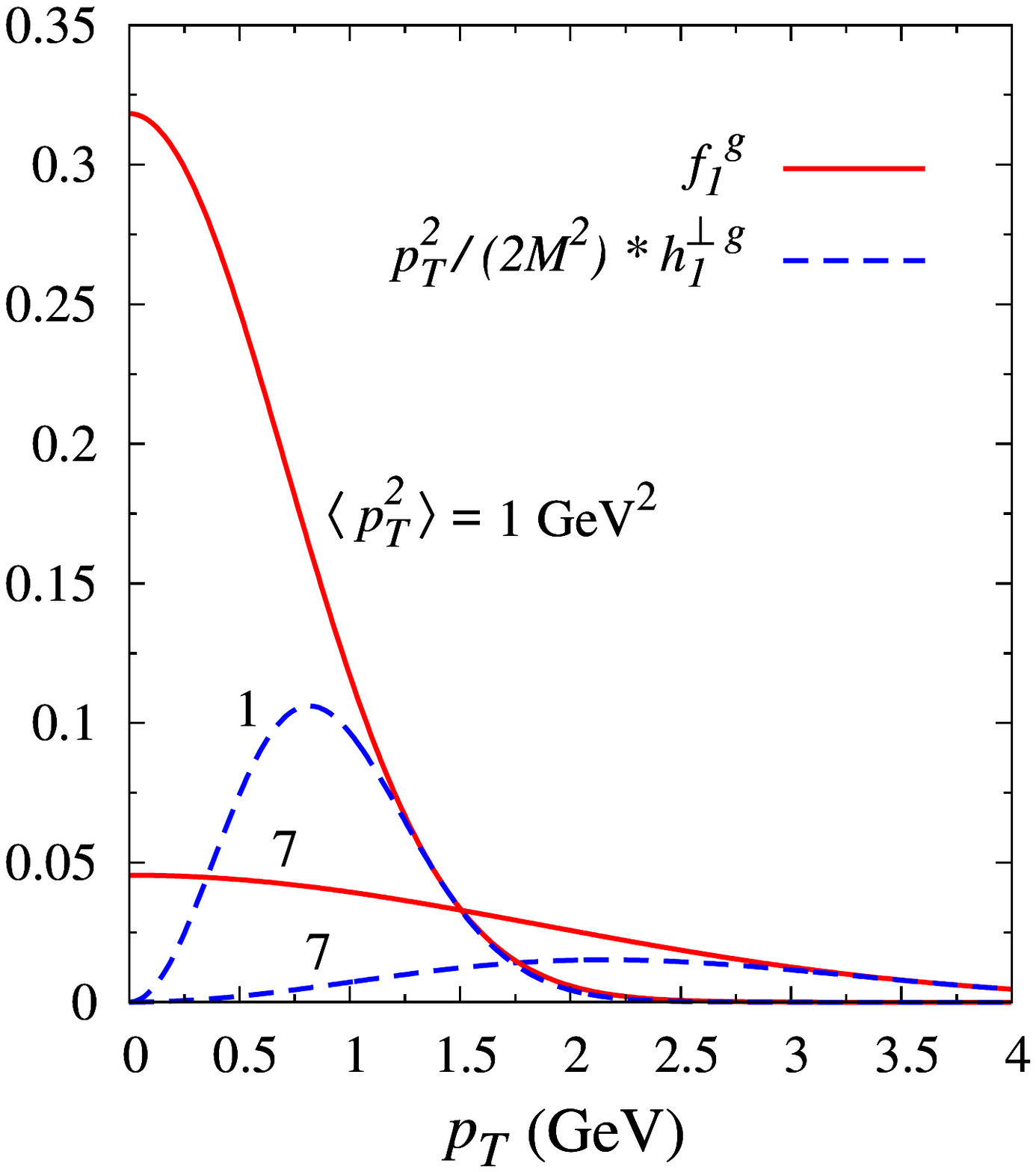, width=0.32\textwidth}

\vspace*{-7.45cm}
\hspace*{3.4cm}
\psfig{file=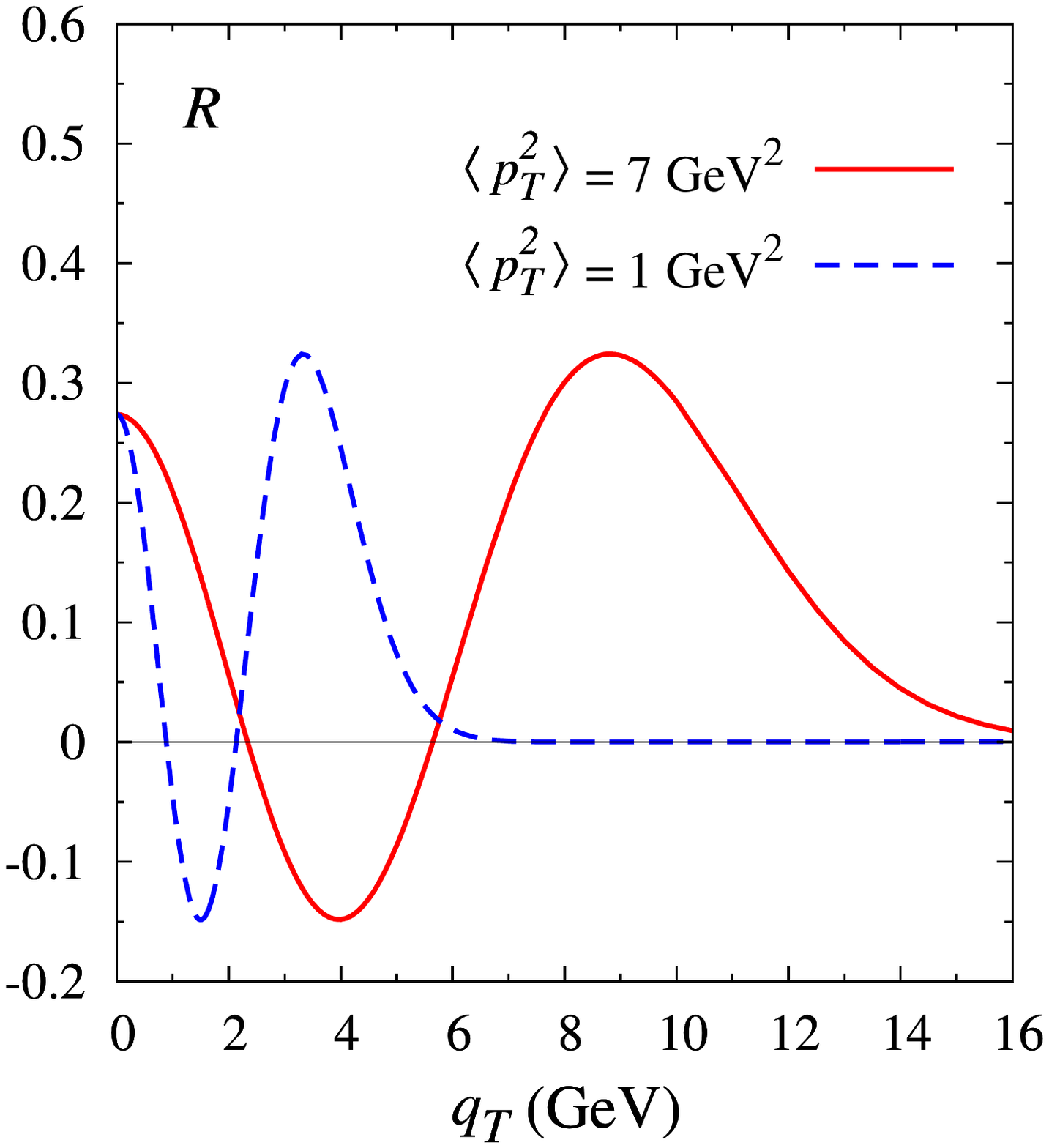, width=0.32\textwidth}

\vspace*{-0.6cm}
\caption{\it Left: Gaussian distributions for $f_1^g$ and $h_1^{\perp \, g}$ (divided by $G(x)$) as functions of $p_\sT$ for two different
values of $\langle p_\sT^2 \rangle$. 
Right: Resulting ratio $R=\mathcal{C}[w_H\,h_1^{\perp g}\,h_1^{\perp g}]/\mathcal{C}[f^g_1\,f_1^g]$.}  
\label{Asym}
\end{figure}

We follow a standard approach for TMDs in the literature (see \cite{Schweitzer:2010tt}) and
assume a simple Gaussian dependence of the gluon TMDs on transverse momentum:
\begin{equation}
f_1^g(x,\bm p_\sT^2) = \frac{G(x)}{\pi \langle  p_\sT^2 \rangle}\,
\exp\left(-\frac{\bm p_\sT^2}{\langle  p_\sT^2 \rangle}\right)\,,\label{eq:Gaussf1}
\end{equation}
where $G(x)$ is the collinear gluon distribution and the width %modDB
$\langle p_\sT^2 \rangle$ is assumed to be independent of $x$. %modDB
The bound (\ref{eq:Bound}) is directly satisfied by the form
\begin{equation}
h_1^{\perp g}(x,\bm p_\sT^2)=\frac{M^2G(x)}{\pi \langle p_\sT^2 \rangle^2}\frac{2e(1-r)}{r}\,
\exp\left(-\frac{\bm p_\sT^2}{r \langle p_\sT^2 \rangle}\right)\,.\label{eq:Gaussh1perp}
\end{equation}
We choose $r=2/3$. 
The left panel of Fig.\ \ref{Asym} shows the $\bm p_\sT$-dependence of $f_1^g$ and $h_1^{\perp g}$
for two values of the Gaussian width: $\langle p_\sT^2 \rangle = 1\,\mathrm{GeV}^2$ and 
$\langle p_\sT^2 \rangle = 7\,\mathrm{GeV}^2$. The latter value may be more appropriate at $Q=m_H$, cf.\ the Gaussian fit 
to $f_1^u(x, \bm p_\sT^2)$ evolved to $Q=M_Z$ of Ref.\ \cite{Aybat:2011zv}. %modDB

%%%%%%%%%%%%%%%%%%%%%%%%%%%%%%%%%%%%%%%%%%%%%%%%%%%%%%%%%%%%%%%%%
%%%%%%%%%%%%%%%%%%%%%%%%%%%%%%%%%%%%%%%%%%%%%%%%%%%%%%%%%%%%%%%%%

It is straightforward to compute the convolution integrals appearing in Eq.~(\ref{eq:CSscalarHiggs}) analytically: %modDB
\begin{align} %modDB
& \mathcal{C}[f^g_1\,f_1^g] =  \frac{G(x_a)\,G(x_b)}{2\pi \langle p_\sT^2 \rangle}\mathrm{exp}
\left(-\frac{\bm q_\sT^2}{2 \langle p_\sT^2 \rangle}\right),\label{eq:GaussConvf1}\\ 
& \mathcal{C}[w_H\,h_1^{\perp g}\,h_1^{\perp g}]  = \frac{G(x_a)\,G(x_b)}{36 \pi 
\langle p_\sT^2 \rangle}\,\times\nonumber\\
& \hspace{0.3 cm} \left[\frac{2}{3} -\frac{\bm q_\sT^2}{\langle p_\sT^2 \rangle}
 +\frac{3(\bm q_\sT^2)^2}{16 \langle p_\sT^2 \rangle^2}\right]\,\mathrm{exp}\left(2-\frac{3 \bm q_\sT^2}{
 4 \langle p_\sT^2 \rangle}\right)\,.\label{eq:GaussConvh1perp}
\end{align}

Their ratio $R$ is shown in the right panel of Fig.\ \ref{Asym}. 
It is a measure of the relative size of the contribution by linearly polarized gluons and shows 
the anticipated double node of $\mathcal{C}[w_H \,h_1^{\perp g} \,h_1^{\perp g}]$. 
It is evident that at least within our simple model, linearly polarized 
gluons have a sizable effect on the Higgs $q_\sT$ distribution. 
We stress again that the effect enters scalar and pseudoscalar Higgs production with opposite sign. If the effect 
is at or near the level shown by Fig.\ \ref{Asym}, it should easily allow to determine the parity of the 
Higgs boson, provided a sufficiently fine scan in $q_\sT$ is possible in experiment, to resolve both nodes.

So far we have considered only the production of an on-shell Higgs boson. In reality the 
Higgs will decay into some observed final state, and there will be background reactions contributing 
to this final state that are not related to the Higgs. These backgrounds may themselves be sensitive
to linearly polarized gluons. We will now briefly consider one example of this, the Higgs decay into a photon 
pair. We reserve a more detailed study of final states such as $\gamma Z$, $ZZ$ or $WW$ for a future publication. 

After production in $gg\to H$, the two-photon decay of a SM Higgs occurs through a top quark or 
$W$-boson triangular loop. The decay of a pseudoscalar Higgs is instead described by physics beyond the 
SM and hence is model-dependent. There are often no tree-level couplings to $W$-bosons in this 
case \cite{Bernreuther:2010uw}, so here we consider only the top quark 
coupling. For both a scalar or pseudoscalar 
Higgs, the lowest-order amplitude can be written as~\cite{Djouadi:2005gi}
\begin{equation}
\hspace{-0.5cm}\hat{\mathcal{M}}_{\gamma\gamma}^{\mu \nu}   = 
\frac{\sqrt{2}G_F \alpha_s \alpha}{32 \pi^2}
\frac{s^2\,{\cal A}_{H(A)}(\bar\tau)\,{\cal F}_{H(A)\to \gamma \gamma}(s)}{s-m_H^2+i\Gamma_H m_H}
r^{\mu \nu}\,,\label{eq:Higgsdecay}
\end{equation}
where $\alpha$ is the electromagnetic coupling and $\Gamma_H$ the Higgs decay width.
For a scalar Higgs, $r^{\mu \nu}=g_\sT^{\mu \nu}\delta_{\lambda_a \lambda_b}$, 
whereas $r^{\mu \nu}=\lambda_a i \epsilon_\sT^{\mu \nu}\delta_{\lambda_a \lambda_b}$ in the pseudoscalar case,
with $\lambda_a,\,\lambda_b$ the photon helicities. 
${\cal A}_{H}(\bar\tau)$ and ${\cal A}_{A}(\bar\tau)$ are given in Eqs.~(\ref{eq:Scalar1}) 
and~(\ref{eq:Pseudo1}) with $\bar\tau=s/(4m_t^2)$, where $s=(p_a+p_b)^2\simeq x_a x_b {S}$ 
for gluon momenta $p_a$, $p_b$. Finally, 
\begin{equation}
{\cal F}_{H \to \gamma \gamma}(s)  = {\cal W}(\tau_W)+\frac{4}{9} N_c  {\cal A}_{H}(\bar\tau) \,,\label{eq:FFs}
\end{equation}
with $\tau_W=s/(4m_W^2)$ and ${\cal W}(\tau)=-(2\tau^2+3\tau+3(2\tau-1)
J(\tau))/\tau^2$ describes the contribution by the $W$ triangular loop.
We assume ${\cal F}_{A\to \gamma \gamma}(s)=\frac{4}{9} N_c {\cal A}_{A}(\bar\tau)$.
In the following we consider a relatively light Higgs mass $m_H=120\,\mathrm{GeV}$ with a small total width $\Gamma_H\simeq 5\times10^{-3}\,\mathrm{GeV}$~\cite{Carena:2002es}.

As is well known, an important QCD background to photon pair production at high energies is 
generated by $gg\to\gamma \gamma$ via a quark box \cite{Dicus:1987fk}. %modDB 
This subprocess was studied recently 
in the context of TMD factorization in Ref.~\cite{Qiu:2011ai}. Using its results, we add the 
two lowest-order amplitudes describing the box diagram and the Higgs resonance and 
extract the azimuthal-angle independent cross section: 
\begin{equation}
\int \d \phi\,\frac{\d \sigma^{gg}}{\d^4q\,\d\Omega}=F_1 \mathcal{C}[f_1^g\,f_1^g]+
F_2\mathcal{C}[w_{H}h_1^{\perp g}\,h_1^{\perp g}].\label{eq:DPCS}
\end{equation}
Here $q=q_a+q_b$ is the momentum of the photon pair. $d\Omega=d\phi\, d\cos\theta$ denotes the solid angle 
element for each photon, with the angles $\phi, \theta$ defined in the Collins-Soper frame \cite{Qiu:2011ai}. $F_1$
and $F_2$ are calculated functions of $\theta$ and the pair mass $Q =\sqrt{s}$ 
that we will not give here. They depend on the box and Higgs amplitudes. 

\begin{figure}[t]
\vspace*{-0.6cm}
\psfig{file=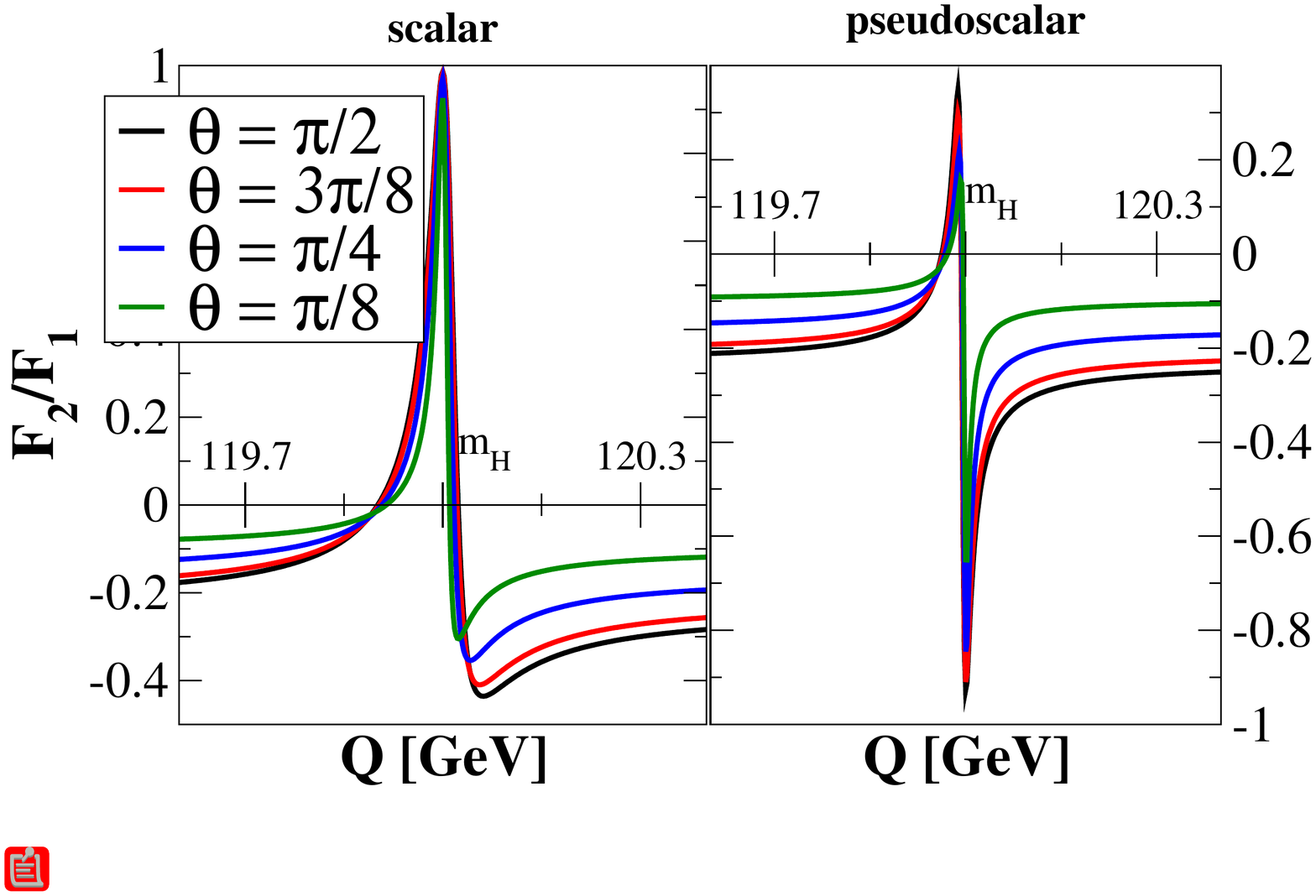,width=0.52\textwidth} 
\vspace*{-2.2cm} 
\caption{\it Ratio $F_2/F_1$ as a function of pair mass squared in a region around 
$m_H=120 \ {\rm GeV}$ for various angles $\theta$.}
\label{fig:F2overF1}
\end{figure}

%In our numerical estimates, %modDB
We find that the box contribution dominates the process 
except when the photon pair mass is close to the Higgs mass. Figure~\ref{fig:F2overF1} shows
the effect of the box-Higgs interference on the ratio $F_2/F_1$ as a function of $Q$ around $m_H=120 \ {\rm GeV}$ %modDB 
for a scalar or pseudoscalar Higgs. Away from $Q=m_H$ (by a few hundred MeV) %modDB
we find $F_1 \gg F_2$, such that the additional term from linearly polarized gluons  
contributes at most 10\% to the cross section, but on average around 1\% or less. %modDB
However, near $Q=m_H$ where the Higgs contribution dominates, we find $F_1 \approx \pm F_2$. %modDB 
The ratio of the second to first term in Eq.\ (\ref{eq:DPCS}) then becomes approximately the ratio $\pm R$ %modDB
of Fig.~\ref{Asym}. Figure~\ref{fig:F2overF1} suggests that a distinction between a scalar and pseudoscalar Higgs is 
possible, if the experimental resolution of the photon pair mass $Q$ is sufficiently good. Higgs bosons in 
extensions of the SM which typically have larger widths, would required less fine $Q$-binning. Also, for heavier Higgs bosons
other final states such as $WW$- or $ZZ$-production may allow for a better $Q$-resolution.
In any case the $Q$-bin size around the Higgs mass is to be chosen as small as possible to maximize the effects caused by linearly polarized gluons. %modDB

We conclude that the effect of linearly polarized gluons on the Higgs transverse momentum distribution 
can in principle be used to determine the parity of the Higgs boson, provided $h_1^{\perp g}$ is of
sufficient size. Of course, it could turn out that $h_1^{\perp g}$ is in reality smaller than in our model or 
that it exhibits nodes in $x$ or $p_\sT$, complicating the analysis. Our results thus provide additional motivation 
for experimental studies of $h_1^{\perp g}$ using different probes, such as dijet and heavy quark or photon 
pair production. We stress that perturbative gluon-radiation effects will alter the $q_\sT$ distributions expected 
on the basis of our simple Gaussian model. Their inclusion will require merging our model with the
soft-gluon resummation techniques described 
in~\cite{Nadolsky:2007ba,Catani:2010pd,Bozzi:2005wk,Aybat:2011zv,Sun:2011iw}. This will also 
affect the eventual size of the contribution by linearly polarized gluons. A full study of this is needed. 

%%%%%%%%%%%%%%%%%%%%%%%%%%%%%%%%%%%%%%%%%%%%%%%%%%%%%%%%%%%%%%%%%%%%%%

%%%%%%%%%%%%%%%%%%%%%%%%%%%%%%%%%%%%%%%%%%%%%%%%%%%%%%%%%%%%%%%%%%%%%%%

\begin{acknowledgments}
We thank John Collins and Feng Yuan for useful discussions.
C.P.~is supported by Regione Autonoma della Sardegna under grant PO Sardegna FSE 2007-2013, L.R. 7/2007.
This research is part of the FP7 EU-programme Hadron Physics (No.\ 227431) and part of the research program of the ``Stichting voor Fundamenteel Onderzoek der Materie (FOM)'' which
is financially supported by the ``Nederlandse Organisatie voor Wetenschappelijk Onderzoek (NWO)''.
W.V.'s work is supported by the U.S. Department of Energy 
(contract DE-AC02-98CH10886).
\end{acknowledgments}
%%%%%%%%%%%%%%%%%%%%%%%%%%%%%%%%%%%%%%%%%%%%%%%%%%%%%%%%%%%%%%%%%%
%%%%%%%%%%%%%%%%%%%%%%%%%%%%%%%%%%%%%%%%%%%%%%%%%%%%%%%%%%%%%%%%%%\

\end{document}